\documentclass[preprint]{elsarticle}

\usepackage{lineno,hyperref}
 \usepackage{amsmath,amssymb}
 
\usepackage[utf8]{inputenc}
\usepackage[T1]{fontenc}
\usepackage[letterpaper,margin=1in]{geometry}
\usepackage{lmodern}
\usepackage{microtype}
\usepackage{graphicx}
\usepackage{dcolumn}
\usepackage{bm}
\usepackage{hyperref}
\usepackage{booktabs}
\usepackage{physics}
\usepackage{siunitx}
\usepackage{isotope}
\usepackage{xspace} 
\usepackage{color}
\usepackage{topcapt}

\modulolinenumbers[1]

\journal{   }  

\begin{document}

\begin{frontmatter}

\title{Quantum Entanglement and Thermal Behavior in Charged-Current Weak Interactions}
\author[mymainaddress]{G. Iskander}
\author[mymainaddress]{J. Pan}
\author[mymainaddress]{M. Tyler}
\author[mysecondaryaddress]{C. Weber}
\author[mymainaddress]{O.K. Baker \corref{mycorrespondingauthor}}
\cortext[mycorrespondingauthor]{Corresponding author \ead{oliver.baker@yale.edu; phone 01-203-432-3374}}

\address[mymainaddress]{Department of Physics, Yale University, New Haven, CT 06520}
\address[mysecondaryaddress]{Physics Department, Brookhaven National Laboratory, Upton, Long Island, NY 11973-5000}

\begin{abstract}
We show that quantum entanglement between causally separated regions of a nucleon 
in antineutrino-nucleon scattering manifests itself as a thermal component in the resulting pion
momentum distribution.  For antineutrino scattering coherently from the (carbon) nucleus as a whole,
this thermal component is absent, as expected by our quantum entanglement thermalization proposition.  These 
phenomena, which have been observed in proton-proton collisions at the Large Hadron Collider, and
in electromagnetic deep inelastic scattering, are now for the first time shown to exist in electroweak 
interactions as well.
\end{abstract}


\end{frontmatter}


\section{\label{sec:level1}Introduction}

The relationship between quantum entanglement, entanglement entropy (EE), and
thermal behavior is of great interest in several subfields of physics.  A non-exhaustive list of
recent examples includes research on quantum entanglement dynamics at
small Bjorken-$x$ in deep inelastic scattering, which is explained as being due to 
quantum entanglement and EE in electromagnetic interactions~\cite{EntanglementEntropy_DIS}.  
The AdS/CFT correspondence has enabled keen insights into EE and its dynamics in 
regions of black holes~\cite{EntanglementEntropy_Blackholes}.
 Quark-antiquark correlations via entanglement entropies in Lattice Gauge Theory 
 are important studies of parton distributions at both the energy and intensity
 frontiers in particle physics~\cite{EntanglementEntropy_LGT}.
Experimental and theoretical investigations in heavy ion and proton-proton 
collisions that include considerations of quantum 
entanglement~\cite{feal_thermal_2019, ho_entanglement_2016} have
provided clarity  and new insights into such phenomena as nuclear 
shadowing~\cite{Entanglement_gluonshadowing} and 
chiral symmetry breaking in nuclear physics~\cite{Entanglement_chiralsymmetry}.

Separate studies of hadron as well as vector and scalar boson dynamics
resulting from the strong interaction in particle collisions have led to explanations of
observable effects at the Large Hadron Collider (LHC). 
The transverse momentum distribution of charged hadrons from particle collisions 
can be decomposed into two components: a ``hard-scattering'' power-law part, and 
a ``thermal'' exponential part~\cite{kharzeev_deep_2017,bylinkin_origin_2014}. The ``hard'' contribution dominates at high transverse momentum ($p_T$) and is well-characterized by quark and gluon scattering. The origin of the thermal components that dominates at lower $p_T$ is less well understood. Recently it has been proposed that the thermal component is a consequence of quantum entanglement between different regions of the colliding protons~\cite{baker_thermal_2018}.  In that study, this emergence of a thermal component was shown to occur also in proton-proton ($pp$) collisions at the Large Hadron Collider. The thermal component described by an exponential fit to the data was shown to not only be present in the transverse momentum distribution of charged hadrons, but also in the $p_T$ spectrum of Higgs bosons resulting from the collisions. Furthermore it was shown
in~\cite{baker_thermal_2018} that in cases where no quantum entanglement is expected, as in diffractive events, the thermal component is absent in the $p_T$ distribution.

In~\cite{kharzeev_deep_2017,bylinkin_origin_2014}, it is argued that the thermal component is a result of entanglement between causally disconnected parts of the nucleon in the interaction. For this reason inelastic $pp$ collisions exhibit a thermal component, while diffractive collisions, where the nucleon as a whole is probed, give rise to only the hard-scattering component, and no thermal behavior. Additionally it is shown in~\cite{kharzeev_deep_2017} that the quantum entanglement between partons can also be probed by deep inelastic lepton scattering.  

If this quantum EE picture of the thermal component’s origin is correct, then it should also manifest itself in neutrino-nucleon interactions where only part of the nucleon is probed.  We show the presence of a thermal component in the momentum distribution of neutral pions from antineutrino - nucleon scattering, while no such component is present in coherent antineutrino - nucleon scattering. The absence of a thermal component in the latter case is due to the antineutrino probing the nucleus as a whole; there is no un-probed region of the nucleus which can be entangled with the probed region.  In this Letter we show evidence, for the first time, that quantum entanglement in electroweak scattering is responsible for the observed phenomena. 

\section{\label{sec:level2} Background and Phenomenology}
\begin{figure}[tp]
\centering
\includegraphics[width=0.8\columnwidth]{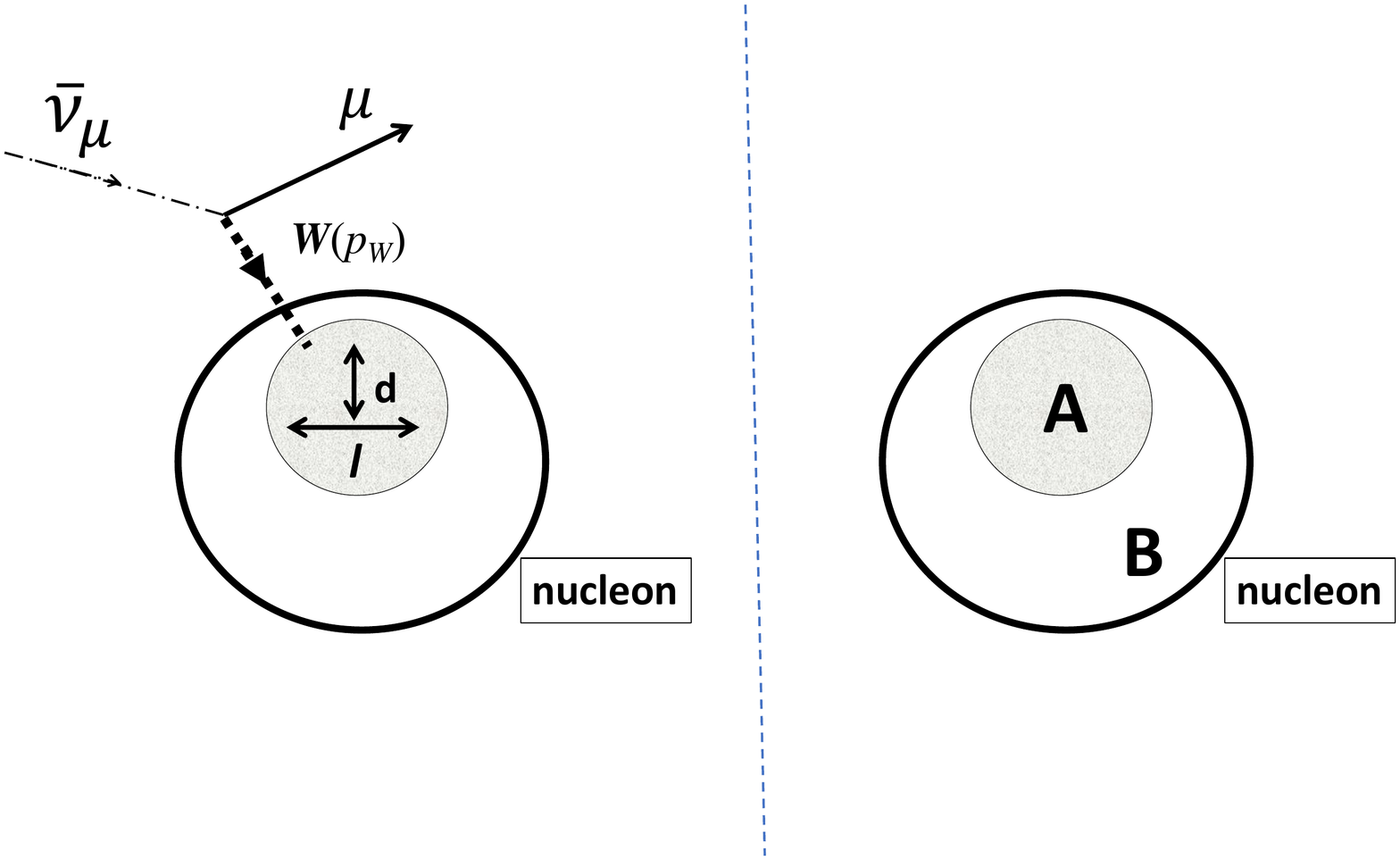}
\caption{(Left side) Antineutrino scattering from a nucleon via emission of a $W$ boson with an exiting muon. The $W$ boson samples a region of the nucleon, not the entire nucleon. (Right side) The region of the nucleon sampled by the interacting $W$ boson is denoted region $A$.  The nucleon region not probed by the boson is region $B$.}
\label{neuscat}
\end{figure}

Charged-current weak interaction processes such as
\begin{equation}
{\bar \nu_{\mu}} + N \rightarrow \mu^+ + \pi^0 + X
\label{eq:one}
\end{equation}
are mediated by $W$ bosons. The vector boson probes only a part of the nucleon wave function, denoted by region $A$ in Fig.~\ref{neuscat}, that has a transverse size of approximately $d = h/p_W$ , and a longitudinal size of approximately $l = (mx)^{-1}$~\cite{kharzeev_deep_2017,bylinkin_origin_2014,baker_thermal_2018}. 
Here $h$ is Planck's constant, $p_W$ is the boson's momentum, $x$ is the momentum fraction carried by the struck quark in the interaction (Bjorken-$x$), and $m$ is the nucleon mass. Within the struck nucleon, the probed region $A$ is complementary to the region $B$ that is not probed in the interaction. The entire space within the nucleon (a pure state) is then $A \cup B$. In the density matrix formalism, the mixed state region $A$ that is probed can be described by 
\begin{equation} 
   \rho_A = \tr_B\rho_{AB} = \sum_n \alpha_n^2 \op{\Psi_n^A}
\end{equation}
\noindent where $\alpha_n^2$ is the probability for existence of a state with 
$n$ partons in this picture and $\tr_B$ is a partial trace with respect to region $B$.  
With this the von Neumann entropy of the state in region
$A$ is
\begin{equation} 
   S_A = - \tr[\rho_A {\rm ln} \rho_A] = - \sum_n \alpha_n^2  \ln \alpha_n^2 
\end{equation}

Intuitively, the von Neumann entropy, also known as the Shannon entropy in classical information theory, is greater in the whole system than in any subsystems. Nonetheless, considering that the whole nucleon $A \cup B$ is in a pure state, its von Neumann entropy
then, by definition, vanishes. 
It was recently proposed in~\cite{kharzeev_deep_2017} that we turn to the literature of quantum information science, where the von Neumann entropy has been extensively studied, to quantify entanglement. 
In the current analysis, as in~\cite{baker_thermal_2018}, the nontrivial $S_A$ is attributed to the quantum entanglement between regions $A$ and $B$ in ~Fig.\ref{neuscat}.
Furthermore $S_A$, evaluated from the QCD evolution equation, is found~\cite{kharzeev_deep_2017} to coincide with the EE calculated in conformal field theory (CFT) in the small-$x$ limit ($x < 10^{-3}$). The correspondence has been discussed in depth in~\cite{kharzeev_deep_2017} and ~\cite{baker_thermal_2018}, among which most notably, the simplicity of the physical interpretation opens up possibilities for broader validity. Since the time scale of the collision $\tau \sim 1/p_W$ is short on the QCD scale, it can be properly mapped to the rapid ``quench'' in the CFT formalism, which generates pairs of entangled states~\cite{kharzeev_deep_2017}.  

The prediction in CFT that carries the most important physical implication is that at late times, $\rho_A$ converges to an exponential distribution, or a thermal state~\cite{Calabrese_2016}. Namely, albeit dis-favored by the conventional mechanism of thermalization, the thermal feature found in the low-$p_T$ region (corresponding to measurement at late times) of the momentum distribution can instead be attributed to the sub-nucleonic entanglement induced by collisions at high energies. 
In this current analysis, we test the hypothesis as briefly captured above in charged-current anti-neutrino interactions at the intensity frontier in particle physics.  We further strengthen our claim by demonstrating that when the nucleus as a whole is scattered by the $W$ boson so that no sub-nucleonic entanglement is produced, the thermal feature is absent from the spectrum, as expected.

\section{\label{sec:level3}Results and Analysis}
The primary process under consideration is neutral pion production in charged-current antineutrino interactions with a CH (hydrocarbon scintillator) target; see Eq.~(\ref{eq:one}). 
The primary analysis considers total inclusive charged current weak interaction differential cross sections, as presented in~\cite{ mcgivern_cross_2016, le_single_2015}.  Ref.~\cite{mcgivern_cross_2016} reports measurements at $\SI{1.5} {\giga\electronvolt} < E_{\nu} < \SI{10} {\giga\electronvolt}$ ($\expval{E_{\nu}} \approx \SI{3.6} {\giga\electronvolt}$) and Ref.~\cite{le_single_2015} reports data at $\expval{E_{\nu}} = \SI{3.6} {\giga\electronvolt}$. The results from both references are presented in this analysis.  Conversion from pion kinetic energy ($T_{\pi}$) in~\cite{mcgivern_cross_2016} to pion momentum in~\cite{le_single_2015} is given by the expression
\begin{equation} 
   \frac{d\sigma}{dp_{\pi}} = \frac{p_{\pi}c^2}{T_{\pi}+m_{\pi}c^2}\frac{d\sigma}{dT_{\pi}}. 
\end{equation}
The relativistic kinetic energy is related to the relativistic energy by
\begin{equation}
    T_{\pi} = m_{\pi}c^2 - m_{0,\pi}c^2
\end{equation}
where $m_{0,\pi}c^2$ is the pion kinetic energy of motion.
Both neutrino and antineutrino cross sections are given versus the (anti)neutrino incident energy.
When the total inclusive neutrino and antineutrino charged current 
$\nu_{\mu} + N \rightarrow \mu^{-} + X$ and $\bar{\nu_{\mu}} + N \rightarrow \mu^{+} + X$ cross sections  are measured; they both show a linear dependence on neutrino energy down to about $\SI{30}{\giga\electronvolt}$.  The antineutrinos show this linear dependence on energy down to $\SI{3}{\giga\electronvolt}$ or below. This kind of behavior is expected for point-like scattering of the antineutrinos from quarks - an assumption which breaks down at lower energies~\cite{particle_data_group_review_2018}.  Hence, this analysis focuses mainly on antineutrino scattering at the energies used in the cross section measurements $(\sim \SI{3.6}{\giga\electronvolt})$. 
We compare the above results against the inclusive charged-current coherent pion production differential cross sections given in~\cite{minera_collaboration_measurement_2018}.

The normalized differential cross section  that is used to describe the thermal behavior from the interaction is given by
\begin{equation} 
   \frac{1}{p_{\pi}} \dv{\sigma}{p_{\pi}}=  A_{\text{thermal}} e^{(-E_{\pi}/T_{\text{thermal}})}
   \label{eq:ThermalBehaviorEq}
\end{equation}
where $p_{\pi}$ ($E_{\pi} = \sqrt{m_{\pi}^2 + p_{\pi}^2}$) is the pion momentum (energy), $A_{\text{thermal}}$ is the normalization factor, and $T_{\text{thermal}}$ is the thermal temperature parameter
\begin{equation}
T_{\text{thermal}} =0.098 \times \qty(\sqrt{s/s_0})^{0.06}\  {\rm GeV},\
\end{equation}
where the Mandelstam variable $s$ is approximately equal to $m_{\text{}}^2 + 2E_{\nu}m_{\text{}}$, and $s_0$ is a normalization constant which is equal to $\SI{1}{\giga\electronvolt}^2$.
The hard-scattering part of the normalized momentum distribution is given by 
\begin{equation} 
  \frac{1}{p_{\pi}} \dv{\sigma}{p_{\pi}} = A_{\text{hard}} \qty(1+ \frac{m_{\pi}^2}{T_{\text{hard}}^2 \cdot n})^{-n} 
  \label{eq:HardBehaviorEq}
\end{equation} 
where $T_{\text{hard}}$ is the hard-scattering parameter 
\begin{equation}
T_{\text{hard}} = 0.409 \times \qty(\sqrt{s/s_0})^{0.06} \text{GeV},\
\end{equation}
$A_{\text{hard}}$ is the relevant normalization factor, and $n$ a power law scaling parameter.  These equations are also discussed in~\cite{baker_thermal_2018,hadro}.

The CERN ROOT fitting program is used to fit these expressions to the MINER$\nu$A results.  A total of five parameters are used in the fitting procedure: $T_{\text{thermal}},\,  T_{\text{hard}},\, n,\, A_{\text{hard}}$,\, and $A_{\text{thermal}}$. In each case, the reduced chi-squared statistic and the fitting parameters with their associated uncertainties are recorded. 
\begin{figure}[h!]
\centering
\includegraphics[width=1.\columnwidth]{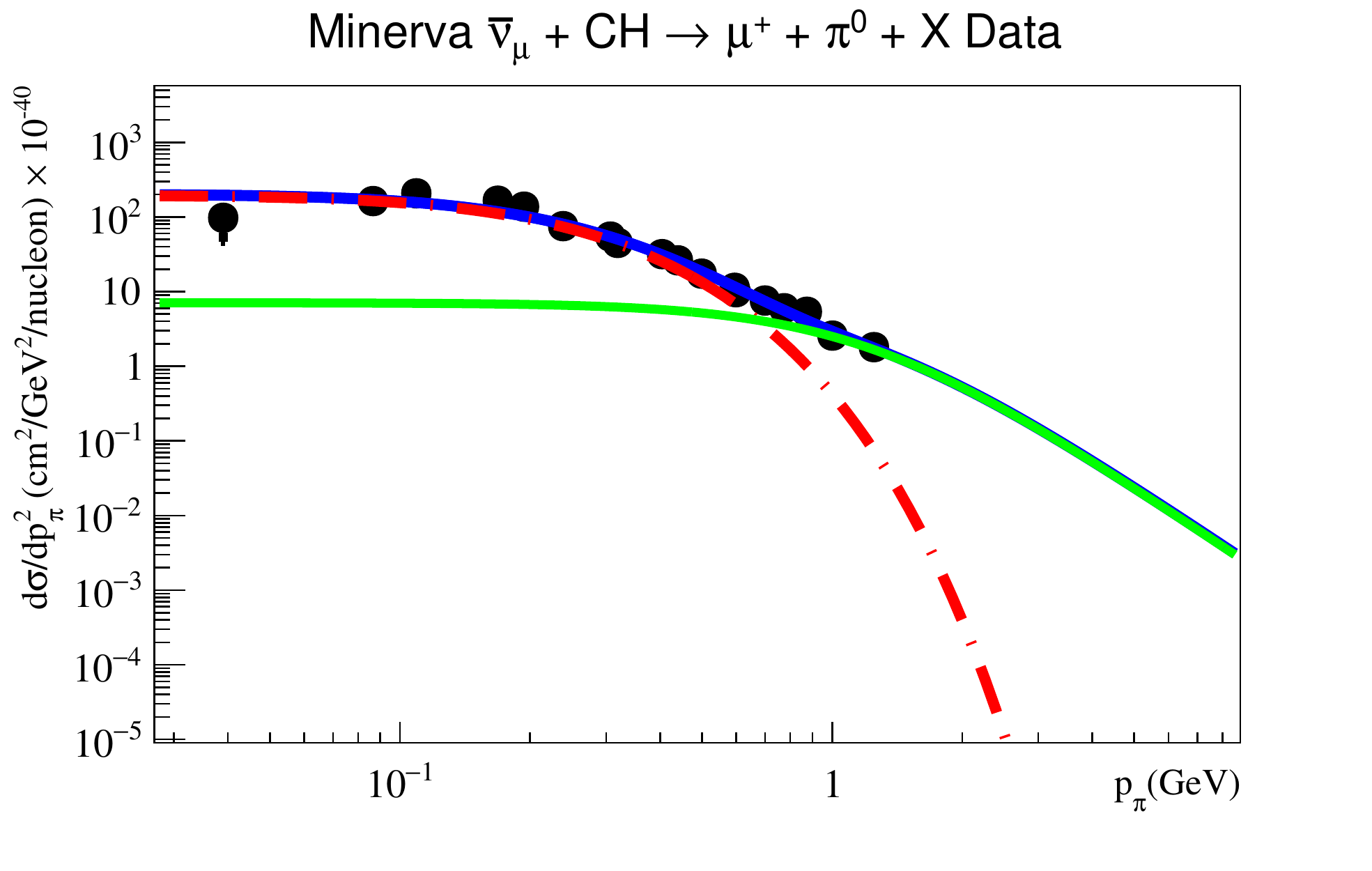}
\caption{Antineutrino differential scattering cross section against hydrocarbon nuclei with charged current pion production.  The dashed (red) line shows the thermal component and the thick solid (green) line shows the hard component of the combined thin solid (blue) line best fit to the data. Data taken from~\cite{mcgivern_cross_2016,le_single_2015}.}
\label{minervapt}
\end{figure}

Fit results of the analysis using data from the MINER$\nu$A collaboration are shown in Fig.~\ref{minervapt}. It is seen from the fit that there are clear and separate thermal (red-dashed)  and hard-scattering (green-full) contributions to the full momentum distributions.  The solid blue curve in Fig.~\ref{minervapt} is the sum of the exponential and power law fits and yields a reduced chi-squared of 
$\chi^2/\text{ndf}  \approx 10.9/13 = 0.84$. Fits of the thermal or hard fit components alone yield $\chi^2/\text{ndf }$ much larger than one.

These results include the effects of final state interactions (FSI). GENIE~\cite{Genie_2010} simulations show that the 
effects of FSI are most significant at low pion momentum values~\cite{mcgivern_cross_2016,le_single_2015}. However, 
given the large statistical and systematic uncertainties at low momenta, we found that the FSI effects did not ultimately 
affect the analysis results or conclusions.

\begin{figure}[!h]
\begin{center}
\includegraphics[width=1.\columnwidth]{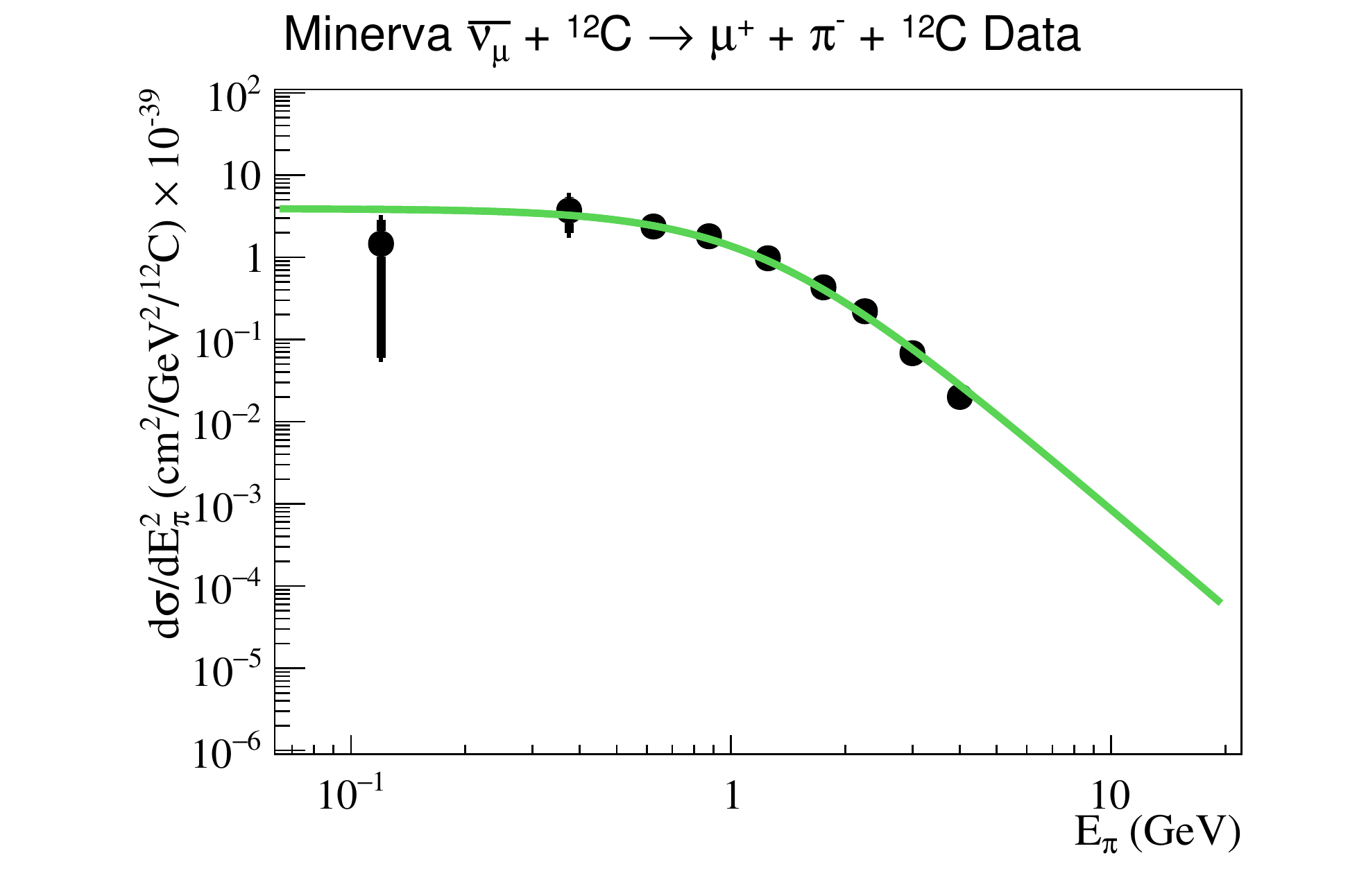}
\caption{Antineutrino coherent scattering from the hydrocarbon scintillator nuclei. The differential cross section is well described by a hard-scattering component (solid green line) alone, as expected in the absence of entanglement. The reduced chi-squared fit is $\chi^2/\text{ndf} = 5.5/6$.  Data taken from~\cite{minera_collaboration_measurement_2018}.}
\label{minervacoherent}
\end{center}
\end{figure}

When the antineutrino in the charged current vector boson interaction scatters coherently from the nucleus $A$ as a whole, as in 
\begin{equation}
\bar{\nu_{\mu}} +  A  \rightarrow  \mu^+  +  \pi^-  +  A ,
\end{equation}
 there is no entanglement expected in the manner described in this paper.  Accordingly, there should be no thermal component in the momentum distribution of the single pion.  Coherent scattering data from the MINER$\nu$A
collaboration~\cite{minera_collaboration_measurement_2018} supports this description of the interaction: The distribution is described well by a power law fit alone, with no thermal component to the distribution.  This is expected due to the absence of entanglement in this process; see Fig.~\ref{minervacoherent}.

\begin{table}[htbp]
   \centering
   \topcaption{$R$ values from different measurements of EE in particle collisions, as defined in Eq.~\ref{eq:RValueDefinition}.} 
   \begin{tabular}{@{} llc @{}} 
      \toprule
      \cmidrule(r){1-2} 
      $R$    & Process & Reference\\
      \midrule
      0.16 $\pm$ 0.05  & $pp$ $\rightarrow$ charged hadrons           & \cite{baker_thermal_2018,weber_quantum_2019}  \\
      0.15 $\pm$ 0.05  & $pp$ $\rightarrow$ H $\rightarrow \ \gamma \gamma$   & \cite{baker_thermal_2018,weber_quantum_2019} \\
      0.23 $\pm$ 0.05  & $pp$ $\rightarrow$ H $\rightarrow \ 4l(e,\mu)$     & \cite{baker_thermal_2018,weber_quantum_2019} \\
      1.00 $\pm$ 0.02  & $pp$($\gamma \gamma) \rightarrow (\mu\mu$) X$'$X$''$   & \cite{baker_thermal_2018,weber_quantum_2019}  \\
   0.13 $\pm$ 0.03  & $\bar{\nu}_{\mu} + N \rightarrow \mu^+ + \pi^0 + X $ &  current work\\
      1.00 $\pm$ 0.05  & $\bar{\nu}_{\mu} + \isotope[12]{C} \rightarrow \mu^+ + \pi^- + \isotope[12]{C} $ &  current work\\
      \bottomrule
   \end{tabular}
   \label{tab:R-tempratio}
\end{table}
The ratio of hard-scattering contributions to the sum of hard and thermal scattering in the distribution, $R$, has been used as an indicator of the presence or absence of quantum entanglement in the interaction \cite{baker_thermal_2018,hadro}.  $R$ is given by the integral (area under the curves) of power law (I$_{\rm p}$) and exponential (I$_{\rm e}$) fits from the figures above:
\begin{equation}
R = \frac{\text{I$_{\rm p}$}}{\text{I$_{\rm p}$} + \text{I$_{\rm e}$}}.
\label{eq:RValueDefinition}
\end{equation}
$R$ is computed from the integral of the combined fit, which combines the hard-scattering function Eq.~\ref{eq:HardBehaviorEq} and the exponential function
Eq.~\ref{eq:ThermalBehaviorEq}. The resulting value of $R$ for the charged-current antineutrino - nucleon interaction is $0.13 \pm 0.03$.  The charged-current coherent antineutrino - nucleus interaction yields an $R$ value of $1.00$. This last value is expected, given the lack of EE and exponential component in the combined fit for this process. The $R$ values obtained in charged-current weak interactions are consistent with values obtained for $pp$ collisions ~\cite{baker_thermal_2018}.  See Table~\ref{tab:R-tempratio} for a listing.

\section{\label{sec:level1}Conclusion}
Our results corroborate those in \cite{kharzeev_deep_2017,bylinkin_origin_2014,baker_thermal_2018,Calabrese_2016}, namely that quantum entanglement in hadrons is responsible for the thermal behavior observed in hadronic collisions. Furthermore, the new results from charged-current neutrino scattering presented here suggest that the thermalization process from entanglement is interaction independent. Present statistical and systematic uncertainties in the MINER$\nu$A data prevent a more comprehensive analysis. This however, emphasizes the need for additional experimental and theoretical studies. The future Brookhaven Electron-Ion Collider will provide deep inelastic scattering measurements at a center of mass energy around $\SI{200}{\ \giga\electronvolt}$~\cite{accardi_electron_2014} which would be valuable for further studies of entanglement entropy. Heavy ion collisions at RHIC and the LHC may also help shed additional light on EE in elliptic flow thermalization and its relation to quantum chromodynamics.

Such studies will grow in importance, especially as alternate hypotheses emerge in independent studies. The ``entropy of ignorance'' discussed in~\cite{duan_entanglement_2020}, for example, might yield a more detailed explanation for the observed thermal behavior. Much work is needed to disentangle the still poorly understood nature of this phenomenon. Particularly the development of additional experimental signatures will be crucial. Recent work has proposed the use of top-antitop quark spin correlations to detect quantum entanglement at the LHC~\cite{afik_enttanglement_2020}. These signatures might open the doors to new studies of entanglement at the energy and intensity frontiers.
Quantum entanglement and sub-nucleon degrees of freedom continue to be an active subfield of study in particle and nuclear 
physics~\cite{tu_einstein-podolsky-rosen_2020,feal_thermal_2019,afik_enttanglement_2020,
robin_entanglement_2020} and beyond.  This present result reported here shows, for the first time, 
a clear indication 
that the thermal behavior observed in the strong, and electromagnetic interactions is also 
manifest in the electroweak interaction.  And in those reactions where no EE expected (coherent
scattering from nuclei), the thermal component is absent in the momentum distribution.  
Entanglement entropy, in the manner described here, is thus exhibited in many phenomena over 
large energy and spatial precision scales.

\section*{Acknowledgments}
JP thanks Meng Cheng and Qing Xia for very helpful discussions.  OKB gratefully acknowledges funding support from the Department of Energy QuantISED Award DE-SC0019592.


\bibliography{refs.bib}

\end{document}